# Co distribution in ferromagnetic rutile Co-doped TiO$_2$ thin films grown by laser ablation on silicon substrates

Nguyen Hoa Hong [1*], Joe Sakai [2], W. Prellier [3], Awatef Hassini [1]

1. Laboratoire LEMA, UMR 6157 CNRS-CEA, Université F. Rabelais, Parc de Grandmont, 37200 Tours, France
2. School of Materials Science, JAIST, Asahidai 1-1, Tatsunokuchi-machi, Ishikawa 923-1292, Japan
3. Laboratoire CRISMAT, UMR 6508 CNRS, ENSICAEN, 6 Bd du Maréchal Juin, 14050 Caen, France

Pure rutile Co-doped TiO$_2$ films were fabricated successfully by the conventional pulsed laser deposition technique on silicon substrates from a ceramic target. Under the right fabrication conditions, Co concentration in the films could be almost the same as in the synthesized target, and films under various conditions all are ferromagnetic well above room temperature. Even though Rutherford backscattering spectroscopy measurements show that Co atoms seem to be mostly localized near the surface of the films and less exist in deeper levels, other experimental evidences show that the ferromagnetism does not come from Co segregations but from the Co-doped TiO$_2$ matrix. Rutile Ti$_{1-x}$Co$_x$O$_2$ thin films grown by a very simple technique on low-price silicon substrates showing Curie temperature ($T_C$) above 400 K appear to be very attractive to applications.

Since the discovery of Matsumoto *et al.* [1] about 2 years ago, Co-doped $TiO_2$ ($Ti_{1-x}Co_xO_2$) thin films have attracted many research groups due to their exhibition of ferromagnetism well above room temperature which is very useful for applications. Growth of this diluted magnetic semiconductor (DMS) by thin film techniques, such as molecular beam epitaxy (MBE) or pulsed laser deposition (PLD)…, provides excellent control of the dopant concentration and the ability to grow single-layered film. However, there are certain issues in this research field at the moment: how to control the concentration of dopant more easily, how to improve the ferromagnetism, and how to clarify the nature of magnetism in those films. So far, Co-doped $TiO_2$ films were deposited from two targets, Ti and Co or $TiO_2$ and Co-doped $TiO_2$ with a very high concentration of Co in order to get very few percents of Co incorporated in the films, by using very sophisticated methods such as combinatorial laser ablation (using the rotation of combinatorial masks), MBE laser ablation, oxygen plasma assisted MBE or co-sputtering.[1-4] Some paper reported about films which were ablated from a ceramic target but it was said that Co did not get into the structure but remained as Co metal [5]. The average magnetic moments per Co atom reported so far are still very modest (as 0.32 $\mu_B$ for laser ablated films [1] and about 1.1 to 1.3 $\mu_B$ for films grown by oxygen plasma assisted MBE [3,4]), and the nature of ferromagnetism was claimed to be caused by Co or cobalt oxide clusters [4-7]. In order to look forward to solving some of those problems, in this work, we have tried to fabricate Co-doped $TiO_2$ films from a ceramic target on silicon substrates by using a conventional PLD system. It is believed that if we use a well-done ceramic target, and control correctly the growth conditions, the films whose dopant concentrations are almost the same as they are in the fixed target with ferromagnetism above room temperature can be obtained.

A polycrystalline target of Co-doped $TiO_2$ with Ti: Co ratio as 0.88 : 0.12 was synthesized by an organic gel-assisted citrate process. The films were deposited by the PLD techniques (248 nm KrF excimer laser, pulses of 5 Hz) on un-etched (100) Si substrates. We applied various conditions: the oxygen partial pressure ($P_{O_2}$) was kept as $1\times10^{-6}$ Torr or $1\times10^{-5}$ Torr, and the energy density was 1.5 J/cm$^2$ or 3 J/cm$^2$. Hereafter, 4 main conditions will be marked as LL (low $P_{O_2}$, low energy density), LH (low $P_{O_2}$, high energy density), HL (high $P_{O_2}$, low energy density) and HH (high $P_{O_2}$, high energy density). The temperature on the substrates was kept as 700$^o$C. After deposition, all films were cooled down slowly to room temperature under the oxygen pressure of 20 mTorr. The typical thickness of the films was 230 nm. The crystalline structure was studied by X-ray diffraction (XRD) with Cu K$\alpha$ radiation ($\lambda$=1.5406Å), using a Seifert for the $\Theta$-2$\Theta$ scan and an X'Pert Philips MRD for the in-plane measurements ($\Phi$-scans). The magnetization measurements were performed by a Quantum Design superconducting quantum interference device (SQUID) system from 0 to 0.5 T in the range of temperature from 400 K down to He temperature, the film morphology was checked by a scanning electron microscope (SEM), and the chemical composition was determined by both energy dispersive X-ray (EDX) and Rutherford backscattering spectroscopy (RBS) methods. The RBS measurements were performed with an incident energy of He$^+$ as of 3.049 MeV, a scattering angle of 170$^o$ and an accumulation charge for each measurement as of 2 µC).

X-ray measurements confirmed that all film are single phased rutile with only rutile peaks appeared in the spectra (for an example, see Fig. 1 for X-ray patterns of the HL film). The films are highly epitaxial with the c-axis of the rutile (around 2.96Å) perpendicular to the substrate plane. Neither Co nor cobalt oxide phase was found in the spectra. Films on Si substrate are mostly c-axis oriented but other diffraction peaks, indexing on the basis of the rutile phase are present, indicating that the film grows with several orientations (probably due to the large lattice mismatch). However, the Φ-scan recorded around the 110 reflection (see the inset of Fig. 1a) shows 90° separated peaks that gives an evidence of in-plane texture of the rutile phase. Similar scans taken on the 220 reflection of Si revealed that the $TiO_2$ rutile layer grows epitaxially, cube-on-cube on Si substrates. In fact, from SEM images (to be discussed later in this report), the HL sample whose X-ray pattern are shown, is the one which has the worst morphology with the presence of some alien parts which are thought to be due to Co segregations but we found no peaks of cobalt or cobalt oxide, and the film is pure rutile. For other better films, the same results are obtained, only rutile peaks appear in the spectra. It is not possible to say very confirmatively that there is no segregation of Co in the films if it is below the detection limit, but it is sure that the Co-doped films on Si are well-established rutile. Although rutile Co-doped films on Si substrates have been already fabricated successfully by co-sputtering from Co and Ti targets [8] but the present study is the first case they are done by a conventional PLD from one ceramic target.

All films with our chosen growth conditions showed ferromagnetic behaviors at room temperature. The magnetization loops are quite similar, except the difference in magnitude of saturated magnetization ($M_s$) and coercivity ($H_C$). The highest $M_s$ achieved in our films was 0.31 $\mu_B$/Co, almost the same as of the $Co_xTi_{1-x}O_2$ film with $x$ = 7% on a $SrTiO_3$ substrate reported by Matsumoto et al.[1]. Fig. 2(a) shows the magnetization as a function of magnetic field taken at 300 K for the LL film. Hysteresis was observed showing that the film is ferromagnetic even at room temperature. $M(T)$ curve taken at 0.2 T in Fig. 2(b) shows that the film has Curie temperature ($T_C$) higher than 400 K.

As mentioned above, a big issue in the field at the moment is how to control the dopant concentration, and to know Co distribution in the films and the nature of ferromagnetism as well. EDX measurements show that 4 films with 4 different conditions have Co content as of 12%, the same as in the synthesized target. SEM images are shown in Fig. 3. The LL and HH films' morphology are similar to each other, rather homogenous among all, but films are full of particles. The surface of LH film seems to be smoothest, even if with few alien particles on it (white parts). The morphology of the HL film is the worst with very big white parts which is believed to be excess Co, CoO or $Co_3O_4$ since those white parts do not have the spherical shape of normal droplets of thin films, but they look like outgrowths. Basically, black and white parts are detected from parts which have different electric conductivities, therefore they are thought to show different compositions. However, we failed to distinguish the difference in compositions of those parts from EDX measurements (The detection limit of EDX technique does not support to specifically determine the composition of nanometer-size-particles). Since seeing neither peaks of cobalt nor cobalt oxides in the XRD does not rule out completely the possibility that excess Co or cobalt oxides exist, then transmission electron microscopy (TEM) measurements must be done in the near future. To know more precisely about the compositions of thin films, RBS measurements were performed. RBS spectra of Co-doped $TiO_2$ films are shown in Fig. 4. Based on the obtained data, the Ti : Co ratio for each case can be estimated to be 90.8 : 9.2 for the LL film, 92.4 : 7.6 for the LH film, 90.0 : 10.0 for the HL film and 93.5 : 6.5 for the HH film. From RBS data, the highest Co concentration is 10% in the HL film whose SEM picture shows some outgrowth of excess Co or cobalt oxide. No reasonable explanation could be given to the SEM pictures of the LL film (with 9.2% Co) and of the HH film (6.5% Co) since they are quite the same and the HH film with lower Co concentration even has some alien particle on it. Then it is not very simple to say that when the amount of Co in the target is large, it gives some excess on the film which leads to Co or cobalt oxide particles/clusters, on the other hand, one must say that the way how Co atoms distribute in the films depends very much on the growth conditions. As seen in Fig. 4, Co atoms were not distributed uniformly in the films: while Ti peaks have simple rectangular shapes, Co peaks have larger height at the right hand side (shallower levels, taken from the surface) and smaller height at the left hand side (deeper levels). Detailed calculations give concrete information, for example, for the LL film, Ti : Co ratio in the depth from 0 up to 40nm is 70 : 30 while in the layer of from 40nm to 230nm-thick, it is 94 : 6, and as the results the averaged ratio of Ti : Co for the film will be 90.8 : 9.2 (as mentioned above). It means that Co atoms are localized mostly near to the surface of the films while they exist less in the deeper levels. This RBS result explains why by EDX the Co content was found to be 12% since signals from atoms near the surface are more sensitive in EDX.

It is known that the saturated magnetization of Co metal is 1.7 $\mu_B$/Co. It was confirmed by the experimental evidence of Co-doped $TiO_2$ films with Co clusters [5]. The value of $M_s$ as 0.31$\mu_B$/Co in our films shows that the ferromagnetism does not come from Co particles or clusters. This is also confirmed by magnetic force microscopy (MFM) measurements: we found no contrast on the surface of the film with LL conditions, for example, or in other words, no particles or clusters were observed and the film is very homogeneous. According to the theory for dopants in Co-doped $TiO_2$ of Sullivan and Erwin [9], it seems that our films were grown in the "rich oxygen condition" and Co dopants were formed primarily in neutral substitutional form, but not interstitial ("poor oxygen condition" makes Co concentrations of substitutional and interstitial Co roughly equal, and $M_s$ must be in between 1 and 2$\mu_B$/Co as in the report of Chamber [4]). It is thought that the magnitude of $M_s$ can be enhanced very much in an appropriate oxygen environment, on the other hand, the homogeneity of the film surely depends strongly on the growth conditions [10].

In conclusion, we have achieved to fabricate rutile Co-doped $TiO_2$ films on silicon substrates by the conventional PLD technique from a ceramic target. Films with various conditions all show ferromagnetism far above room temperature. Co concentration in the film can be almost the same as in the synthesized target if we apply suitable growth conditions. Even though the distribution of Co is not uniform with Co atoms lie mostly near to the surface of the films, the ferromagnetism in our Co-TiO$_2$ films seemingly does not come from Co metals or clusters. Co-TiO$_2$ films with very high $T_C$ (above 400 K) fabricated by a very simple technique on low-price silicon substrates are useful for applications. However, the magnitude of saturated magnetization is still modest and a higher homogeneity is still looked forward. Improvements can be done by changing the concentration of dopant in the target and adjusting growth conditions.

**Acknowledgements**

The authors would like to thank Dr. A. Ruyter for MFM measurements.

Electronic mail: hoahong@delphi.phys.univ-tours.fr

**Figure Captions**

1. X-Ray diffraction pattern for a Ti $_{0.908}$Co$_{0.092}$O$_2$ thin film deposited on a Si substrate under the oxygen pressure of $1\times10^{-6}$ Torr and the fluence of 3J/cm$^2$. The bump around $2\Theta = 40°$ results from various diffraction peaks arising from different out-of-plane orientations of the TiO$_2$ phase. Only diffraction peaks corresponding to the rutile phase were observed. The inset depicts the $\Phi$-scan recorded for the (110) reflection of the rutile TiO$_2$.

2. Magnetization of a Ti $_{0.908}$Co$_{0.092}$O$_2$ thin film deposited on a Si substrate under the oxygen pressure of $1\times10^{-6}$ Torr and the fluence of 1.5J/cm$^2$ (a) versus magnetic field at 300 K and (b) versus temperature under 0.2 T.

3. SEM images of Co-doped TiO$_2$ films with 4 different conditions: a) LL (P$_{O_2}$ of $1\times10^{-6}$ Torr and fluence of 1.5J/cm$^2$, b) LH ($1\times10^{-6}$ Torr, 3J/cm$^2$, c) HL ($1\times10^{-5}$ Torr, 1.5J/cm$^2$ and d) HH ($1\times10^{-5}$ Torr, 3J/cm$^2$).

4. RBS spectra of Co-doped TiO$_2$ films with 4 different conditions: a) LL, b) LH, c) HL and d) HH.

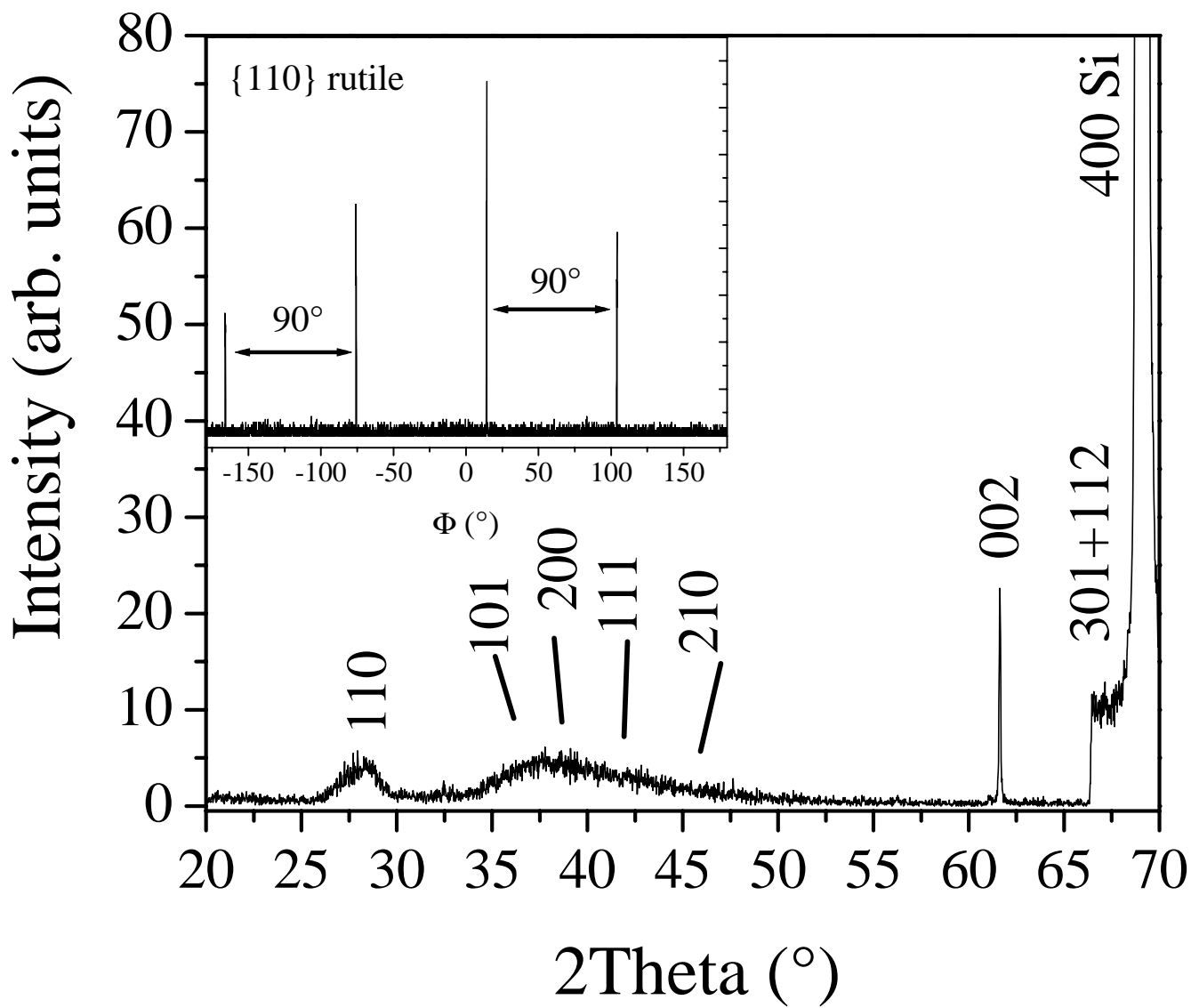

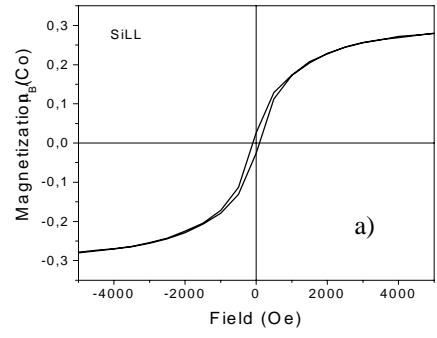

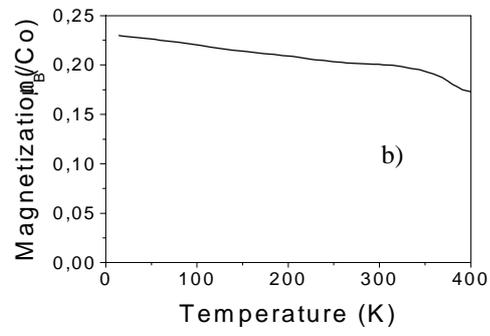

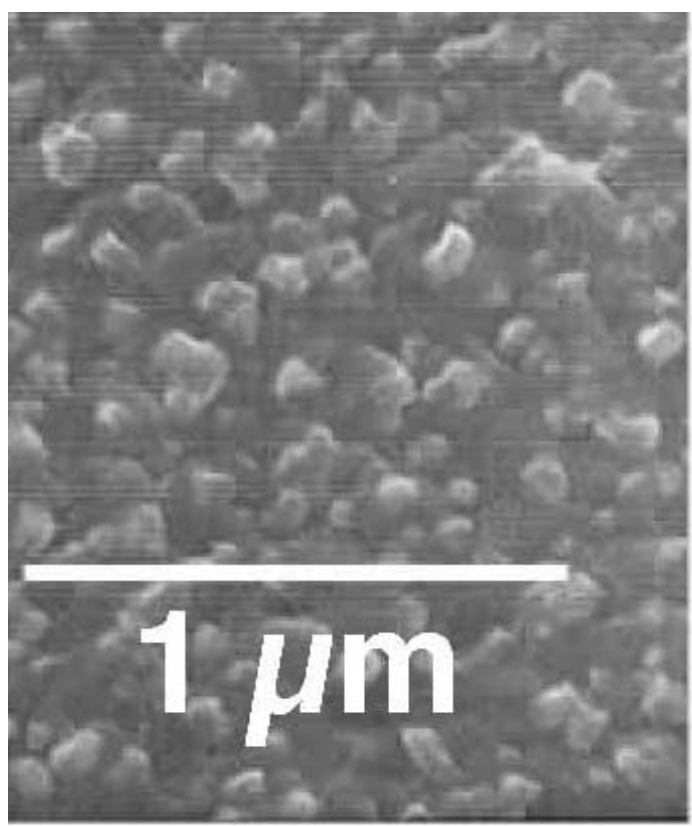
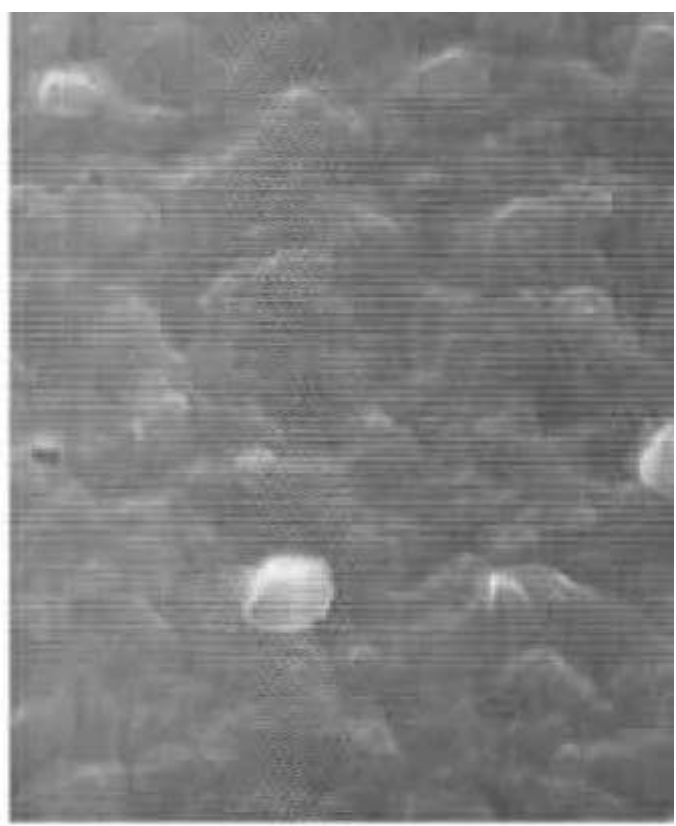
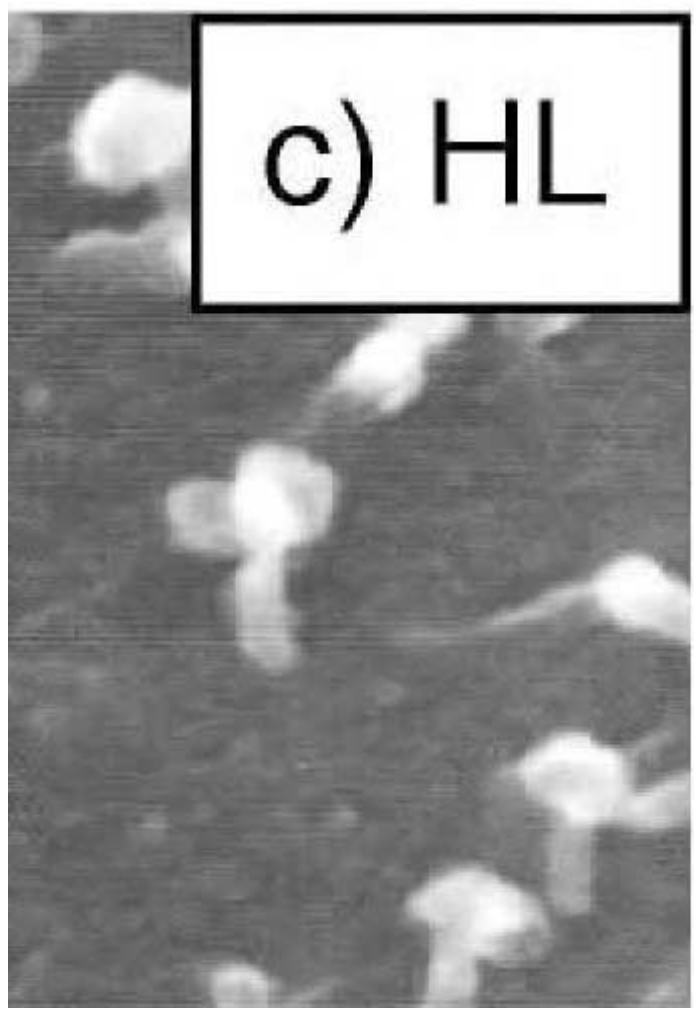
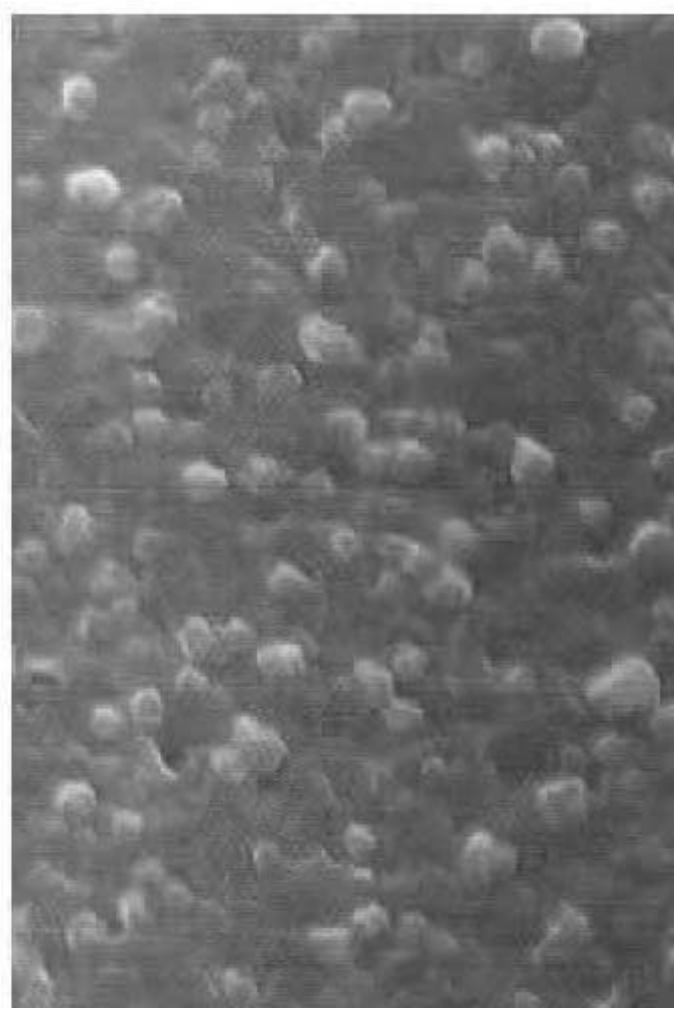

c) HL

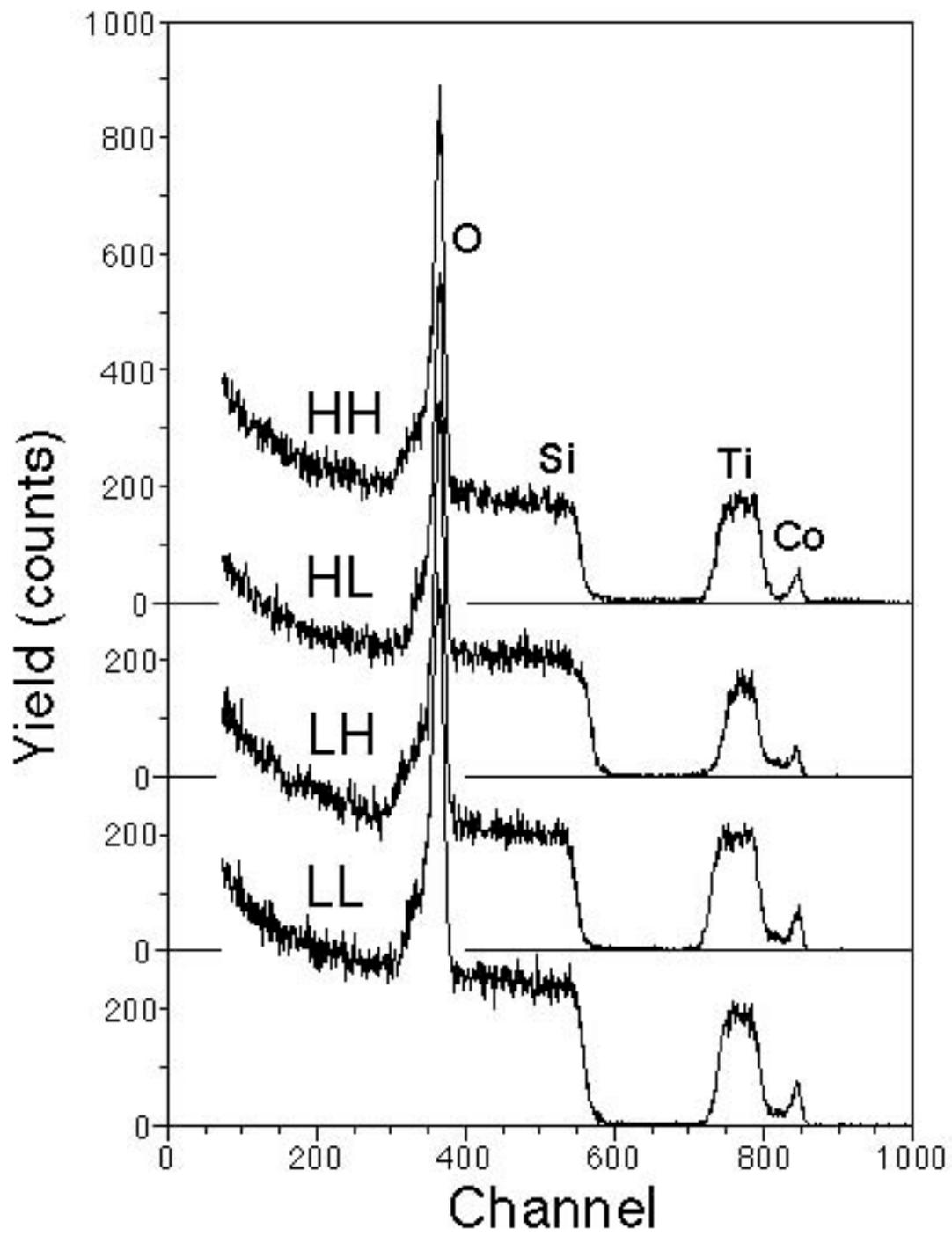